# Core-shell microgels via precipitation polymerization: computer simulations

*Alexey A. Gavrilov*[*,§], *V.Yu. Rudyak*[§], *A.V. Chertovich*[§,#]

[§]Physics Department, Lomonosov Moscow State University, Moscow 119991, Russian Federation

[#] Semenov Federal Research Center for Chemical Physics, Moscow 119991, Russian Federation

**Corresponding Author**

*e-mail: gavrilov@polly.phys.msu.ru

ABSTRACT In this work we presented a novel computational model of precipitation polymerization allowing one to obtain core-shell microgels via a realistic cross-linking process based on the experimental procedure. We showed that the cross-linker–monomer reactivity ratios $r$ are responsible for the microgel internal structure. Values of $r$ lower than 1 correspond to the case when alternating sequences occur at the early reaction stages; this leads to the formation of microgels with pronounced core-shell structure. The distribution of dangling ends for small values of $r$ becomes bimodal with two well-distinguished peaks, which correspond to the core (short dangling ends) and corona (long dangling ends) regions. The density profiles confirm the existence of two distinct regions for small $r$: a densely cross-linked core and a loose corona entirely consisting of dangling ends with no cross-linker. The consumption of the cross-linker in the course in the microgel formation was found to be in a perfect agreement with the predictions of Monte Carlo (MC) model in the sequence space.

Microgels are cross-linked polymer systems of sub-micron size with high tunability in softness, permeability deformability. The importance of microgels for many different applications relies on their high sensitivity to external stimuli (temperature, pH, solution composition, etc) and their low response times. This makes microgels one of the most attractive soft matter objects currently.[1,2] Microgels could be used as delivery carriers,[3,4] stimuli-responsive stabilizers of emulsions,[5–8] photonic crystals,[9] membrane materials[10] and many other technologies[2,11,12]. One of the most widely used microgels, poly(N-isopropylacrylamide) microgels (PNIPAM) are synthesized by relatively simple surfactant-free precipitation polymerization process. During this process, individual chains collapse in poor solvent during growth and form united network during copolymerization of monomers with a small fraction of cross-linker without any additional surfactant.[13] In spite of simplicity of the experimental implementation of this process, its kinetics and, consequently, the internal structure of the resulting microgel particles are quite complex.

Various coarse grained computer models were developed to address this issue.[14] Nikolov et al.[15] prepared microgel particles by randomly distributing cross-linkers within a cubic simulation box and then connecting neighbour cross-linker particles by subchains. Gnan et al.[16] obtained random microgels by self-assembly of patchy particles mimicking the polymerization process of PNIPAM. The resulting structures generally showed more realistic properties compared to regular lattice model.[17] However, the polymerization kinetics in this model is completely unlike typical radical polymerization process due to the patchy particle step-growth mechanism. Thus, it does not reproduce high sensitivity of real precipitation polymerization process to concentrations of initial components, solvent quality, and the properties of the chemical reaction involved. As a result, this model required specific unphysical tweaks to obtain non-homogeneous microgels.[18]

Recently, we proposed a model[19] which mimics the realistic radical-free precipitation polymerization of microgels by applying "mesoscale chemistry" concept to implicit solvent coarse grained molecular dynamics. It allowed to simulate the precipitation polymerization process from a dilute solution of initial components to a final microgel particles up to 70 nm in diameter. The behavior of the simulated microgels was found to be in good agreement with the experimental PNIPAM systems in terms of synthesis kinetics, resulting structure factor and collapse–decollapse curves.

In this paper, we further investigated the model of microgel precipitation polymerization process. We studied the influence of the ratio between reaction rates on the final structure and demonstrated that our model allows natural obtaining both homogeneous and core–shell structures by varying reaction rates of monomer–monomer and monomer–cross-linker.

In order to simulate the precipitation polymerization of a microgel particle, we used the approach recently used in the work [19]; this method follows the logic of the experimental synthesis and does not incorporate any artificial processes. Classical coarse-grained molecular dynamics (MD) with Lennard-Jones potential is used for simulations; to simulate the poor solvent conditions, the potential parameters were the following for all particles: $\varepsilon=2$, $\sigma=1.0$, $R_{cut}=1.4$. The chemical reactions are simulated using the widely used Monte-Carlo simulation approach.[20] Basically, radical polymerization is simulated; for simplicity, we do not take into account the chain transfer and termination processes, so only the initiation and chain growth (propagation) are simulated. Initially, the system contains initiator (i), monomer (m) and cross-linker particles (c). The total number of particles in our simulations was equal to 50000, the ratio of components i:c:m was equal to 0.5:2.5:97 (i.e. 5 cross-linkers per each initiator). The cross-linking mechanism is presented in the Supporting Information.

In our simulations we utilize the terminal chain growth model, which means that the probability of the bond formation depends on the type of the growing end and the particle the growing end attempts to form bond with. The reaction matrix is defined by the following probabilities: $p_{im}$, $p_{ic}$ (these are the initiation probabilities), $p_{mc}$, $p_{mm}$, $p_{cm}$, $p_{cc}$. The latter four govern the resulting reaction sequence (we assumed that the probability of the reactions including cross-linker is independent of its state); if the reactivity ratios $r_m = p_{mc}/p_{mm}$ and $r_c = p_{cm}/p_{cc}$ are small, then the cross-reactions between the monomers and cross-linkers are predominant in the system. This situation leads to a high cross-linker consumption rate at the early stages of the synthesis process, which presumably results in the highly inhomogeneous structure of the resulting microgel since only monomers are present in the system at the late stages of the synthesis process.

It is known from the literature data[21] that in typical experimental systems such as NIPAM/BIS (N, N′-methylenebisacryl-amide) the cross-linker–cross-linker reaction rate for a is two orders of magnitude lower than that for the cross-linker–momomer and monomer–monomer reactions, so the former reaction was neglected in the present work, and the probabilities of the cross-linker–momomer and monomer–cross-linker reactions were assumed to be equal, $p_{mc} = p_{cm}$. Therefore, the synthesis process is controlled by the reactivity ratio $r_m$; in what follows, this parameter will be referred to as $r$. It is worth noting that even in the case of $p_{cc} = p_{mm}$ the reactivity ratio $r_c$ (varied from 0 to the corresponding value of $r_m$) was not found to have any influence on the resulting microgels simply because of the small cross-linker concentration in the system.

Snapshots of the microgels obtained at different values of r are presented in Fig.1.

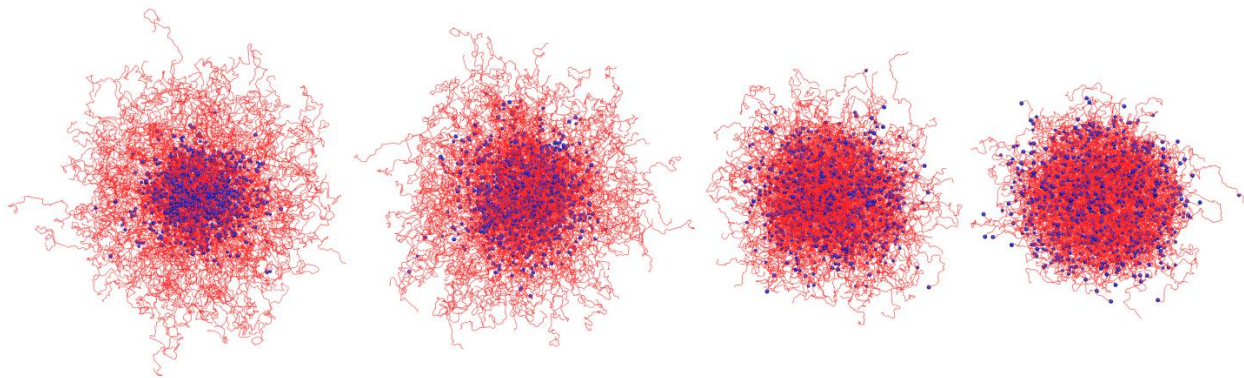

Fig.1 Typical snapshots of the swollen microgels obtained at (from left to right): $r=0.1$, 0.2, 0.5 and 1. The cross-linker particles are presented as blue spheres, whereas the monomer units as red lines. The scale of all images is the same.

We immediately see a striking difference between the microgels: for the large values of r more or less homogeneous structures are observes, while for the small values the cross-linker is mainly located in the microgel center, and, therefore, the microgel is divided into two very well pronounced regions: cross-linked core and loose corona. We will show that these regions indeed are distinctly different.

First of all, let us study the resulting network topology. The most obvious way to do that is to investigate the subchain (i.e. linear chain between two cross-linkers) length distributions; the resulting dependencies of the subchain fraction on their length for different $r$ values are presented in Fig.2(top). Four microgels were synthesized for $r$=0.2-1, whereas for $r$=0.1 we generated 6 microgels for better averaging.

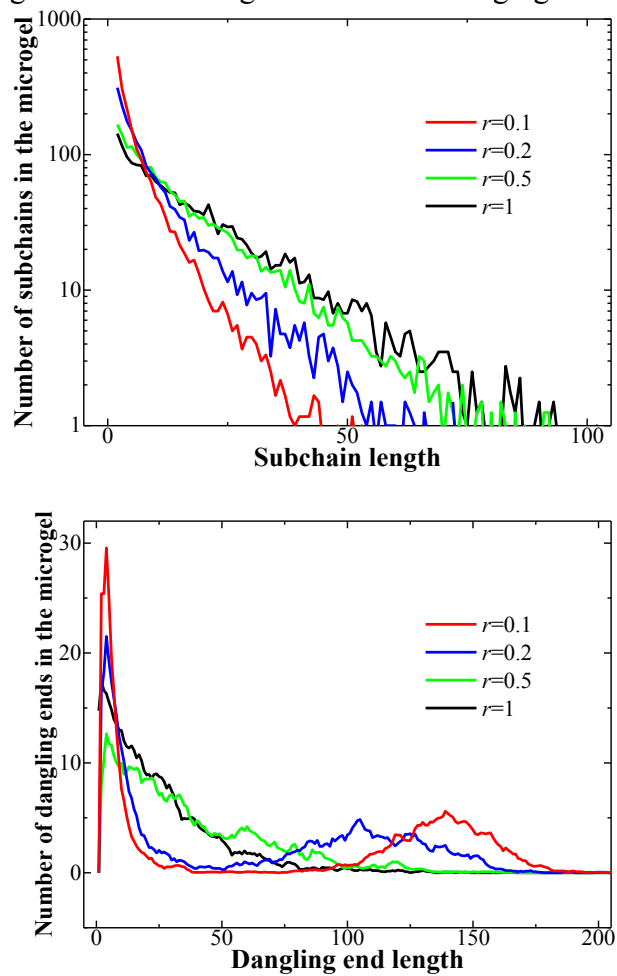

Fig.2 Dependences of the subchain length (top) and the dangling ends length (bottom) in the obtained microgels for different values of $r$. Simple cycles (i.e. formed by a single cross-linker within one linear chain) were excluded from the subchain length distribution. The dangling ends length distribution was calculated using the bin size of 5 in order to reduce the noise.

We see that for $r$=1 and 0.5 the subchains have exponential length distribution; however, upon further decreasing $r$ to 0.2 and 0.1 the curve shape becomes pronouncedly non-linear. This happens due to the fact that at different reaction stages the fraction of available cross-linker in respect to free monomers changes: the majority of the cross-linking reactions occur at the early stages due to $r<1$ (resulting in many short subchains), while at the late stages mainly long subchains are formed since there is only small amount of free cross-linkers available. The average subchain length, obviously, significantly decreases, being equal to 19.1 at $r=1$ and 6.7 at $r=0.1$. The distributions of the dangling ends give us another confirmation of the uneven consumption of the cross-linkers throughout the reaction process; such distributions are presented in Fig.2 (bottom). We can see that the distributions for $r=0.2$ and 0.1 are in fact bimodal: there is a significant number of short and long dangling ends, but almost no intermediate ones. This can easily be understood if we remember (Fig.1) that for these values of r we observe pronounced core-corona structures; therefore, the short dangling ends are mainly found in the core region, whereas the long ends belong to the corona region. For larger values of $r=0.5$ and 1 we see no additional peak as the size of the corona becomes very small, i.e. the microgel is almost homogeneous. It is worth noting that the subchain distributions do not show any bimodality even at the smallest studied value of $r=0.1$, which is in fact obvious given that there are (almost) no subchains in the corona but only dangling ends. It is also worth noting that the average mass fraction of all dangling ends in the microgel significantly increases upon decreasing r, from 24% to 68% at $r=1$ and 0.1, correspondingly.

Next, we studied the density distributions for the microgels obtained for different values of $r$; Fig.3 (top) demonstrates such distributions for the minimal and maximal studied values of $r$, i.e 0.1 and 1.

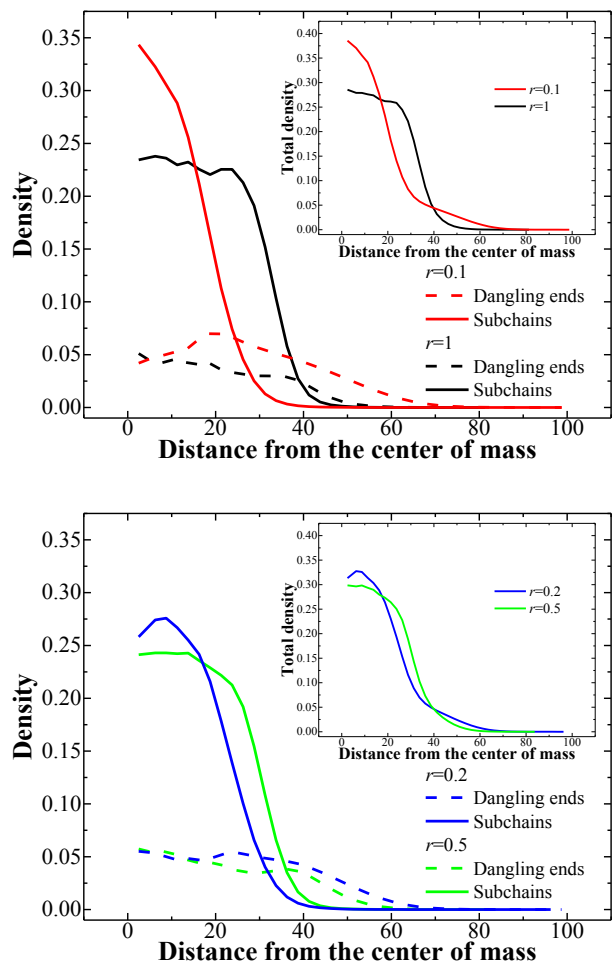

Fig.3 Density profiles of the obtained microgels obtained at different values of $r$: 0.1 and 1 (top) and 0.2 and 0.5 (bottom). The density of subchains and dangling ends are shown separately. The insets shows the total density (i.e. subchains+dangling ends) profiles.

We see that both the subchains and dangling ends in the microgel obtained at $r=1$ are almost homogeneously distributed within the microgel volume, which is to be expected and in good agreement with the results of the previous work.[19] A very small region can be seen at $d>40$ where there are only dangling ends present , but the density in that region is very low, so it is formed by individual randomly oriented outwards dangling ends and cannot be considered as a distinguished corona. On the contrary, for the microgel obtained at $r=0.1$ the density of subchains (as well as total density) rapidly drops when approaching the microgel periphery. The density of the dangling ends demonstrate a maximum at a nonzero distance; moreover, there is a large region at $d>30$ where only dangling ends are present. Therefore, such microgel indeed has two distinct regions: a densely cross-linked core with a small amount of dangling ends and a loose corona entirely consisting of dangling ends with no cross-linker. Moreover, the presence of these two regions is apparent even from the total density profiles: indeed, for $r=0.1$ we see a shoulder at distances from the center of mass of 40-60, which is not present for the case of $r=1$.

We see that for $r$=1 the density profile can almost perfectly be fitted by the model with sigmoidal density decay, see Supporting Infromation, Fig. S1; the linear decay of the density in the core region has been reported for microgels with high enough cross-linker content.[22] For $r$=0.1 the density profile is more complex due to the presence of the second shoulder associated with the corona region. We therefore can speculate that the reason behind the discrepancies between the experimental data and the predictions of the Fuzzy Sphere model[23] (including scattering profiles, see Fig.S1) can be related to the more complex density distribution for core-shell microgels obtained for small enough reactivity ratios $r$.

The density profiles for the intermediate values of $r$=0.2 and 0.5 are presented in Fig.3, bottom. We see that for $r$=0.5 the microgel has similar density profiles to that obtained at $r$=1; the subchain density, however, is constant only in a small volume close to the microgel center. The region where only dangling ends present at $d$>40 is noticeably larger than for $r$=1, but the size of that region and its mass is still rather small; the total density distribution does not have an additional shoulder corresponding to the corona region. For the microgel obtained at $r$=0.2 there is, on the contrary, a pronounced core-shell structure with corona having roughly the same size as the core. The total density distribution also shows the presence of a shoulder similar to the case of $r$=0.1. These observations are consistent with the analysis of the distributions of the dangling ends length (Fig.2, bottom): the distribution is bimodal for $r$=0.1 and 0.2 indicating the presence of two distinct regions in the microgel, while for the larger values of $r$=0.5 and 1 there is no well distinguished second peak responsible for the corona region.

The microgel synthesis process, involving two types of reagents (monomers and cross-linkers), can be viewed as a copolymerization process of two monomers. As it was mentioned above, in our simulations we utilize the classical terminal model of the growth process; we can compare the monomer and cross-linker consumption rate observed in the simulations with the data obtained using Monte Carlo (MC) model in the sequence space.[20] In short, this approach does not take into account the distribution of the monomers and cross-linker in space, but rather calculates the copolymerized sequence using the probabilistic approach based on the remaining concentration of reagents and their reactivity ratios $r$; more detailed description can be found in the work [20]. The dependencies of the fraction of cross-linker consumed (i.e. having reacted at least once) in the course of the reaction on the conversion degree for different values of r are presented in Fig.4.

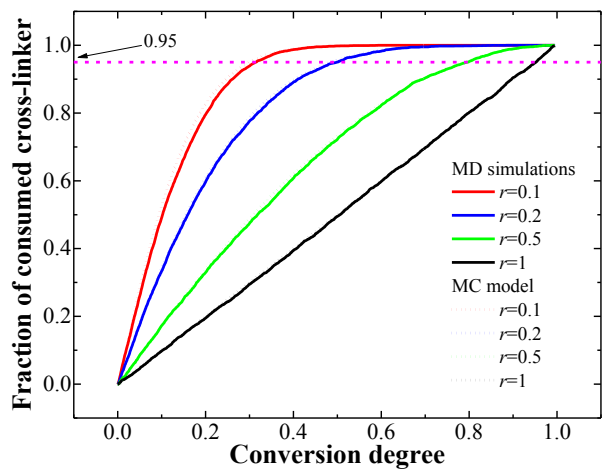

Fig.4 Dependencies of the fraction of cross-linker consumed in the course of the reaction on the conversion degree for different values of r obtained in MD simulations in comparison to the predictions of MC model.

Surprisingly, we see a perfect agreement between the MD simulations data and the results obtained using the MC model despite the presence of phase separation (i.e. the growth of the microgel itself) and polydispersity in the former.

These dependencies could be used to estimate the mass fraction of the corona. The total mass of all dangling ends belonging to the well-separated second peak of the corresponding distribution (Fig.2, right) gives us the actual corona mass for $r=0.1$ and $0.2$; for $r=0.5$ we calculated the mass of all dangling ends longer than 50, which corresponds to a weak local minimum of the distribution. The mass fractions of the corona obtained this way are the following: 0.7 for $r=0.1$, 0.52 for $r=0.2$ and 0.2 for $r=0.5$. If we now look at Fig.4 we see that these values correspond to fraction of the consumed initiator of 95%. These findings can be rationalized in the following way: initially, our system contained 5 cross-linkers per each growing end, and 5% of this value means that there are 0.25 cross-linkers per each growing end left. At such cross-linker content even in the ideal case when each cross-linker connects two different chains, 50% of the ends are not cross-linked, which indicates that the dangling ends dominate in the system after that point. Therefore, we can speculate that the predictions of a fairly simple MC kinetic model can give one a rough estimate of the mass fraction of the corona if the reactivity ratios of the reacting species are known.

Summarizing, in this work we presented a novel computational model of precipitation polymerization allowing one to obtain core-shell microgels via a realistic cross-linking process based on the experimental procedure. In this model, the reaction is simulated by the widely used Monte-Carlo polymerization process mimicking radical polymerization within the molecular dynamics method, and the formation of the microgel particle occurs through the precipitation of large enough polymers formed in the reaction course; no artificial constraints or processes are present in our model (thus mimicking experimental conditions), making it easy to utilize and the results readily comparable with the experimental data. We showed that the cross-linker–monomer reactivity ratios $r$ are responsible for the microgel internal structure. Values of $r$ lower than 1 correspond to the case when alternating sequences occur at the early reaction stages; this leads to high cross-linker consumption, and, therefore, the formation of a densely-cross-linked region (core). At the late reaction stages there is no cross-linker left in the system, and a loose corona is formed consisting of dangling ends; this behavior confirms the previous hypotheses concerning the formation of the core-shell microgels.[24–26] We showed that the network topology dramatically changes upon decreasing the value of $r$; while for r=1 the distribution of subchains is predictably exponential, for smaller values of $r$ it is not the case due to the different cross-linker content in the reaction mixture at different reaction stages. The distribution of dangling ends reveals an important feature: for small values of $r$=0.1 and 0.2 it becomes bimodal with two well-distinguished peaks, which correspond to the core (short dangling ends) and corona (long dangling ends) regions. The density profiles confirms the existence of two distinct regions for $r$=0.1 and 0.2, a densely cross-linked core with a small amount of dangling ends and a loose corona entirely consisting of dangling ends with no cross-linker, while for $r$=0.5 and 1 the structure was homogeneous. The consumption of the cross-linker in the course in the microgel formation was found to be in a perfect agreement with the predictions of Monte Carlo (MC) model in the sequence space. We speculate that such model can be used to estimate the mass of the corona by calculating the remaining monomer mass when the cross-linker content becomes too low to cross-link more than 50% growing ends.

AUTHOR INFORMATION

**Notes**

The authors declare no competing financial interests.

ACKNOWLEDGMENT

The research is carried out using the equipment of the shared research facilities of HPC computing resources at Lomonosov Moscow State University. The work was supported by RSF project 17-73-20167.